\documentclass[prd,tightenlines,nofootinbib,showpacs,preprintnumbers,superscriptaddress]{revtex4}
\usepackage{amsfonts,amsmath,amssymb,amsthm,bbm,hyperref}
\usepackage{graphicx}
\usepackage{color}

\newcommand{\R}{{\mathbb R}}

\newcommand{\cH}{{\mathcal H}}

\newcommand{\hh}{{\mathbf h}}
\newcommand{\cM}{{\mathcal M}}
\newcommand{\cN}{{\mathcal N}}

\newcommand{\be}{\begin{equation}}
\newcommand{\ee}{\end{equation}}
\newcommand{\beq}{\begin{eqnarray}}
\newcommand{\eeq}{\end{eqnarray}}

\newcommand{\f}{\frac}
\newcommand{\vphi}{\varphi}
\def\pp{\partial}
\def\dagg{^\dagger}

\begin{document}

\title{Effects of a scalar field on the thermodynamics of interuniversal entanglement}
\author{I\~naki Garay}
\affiliation{Programa de P\'os-Gradua\c{c}\~ao em F\'isica, Universidade Federal do Par\'a, 66075-110, Bel\'em, PA, Brazil.}

\author{Salvador Robles-P\'{e}rez}
\affiliation{Departamento de F\'{\i}sica Te\'orica, Universidad del Pa\'{\i}s Vasco, Apartado 644, 48080 Bilbao, Spain.}
\affiliation{Estaci\'{o}n Ecol\'{o}gica de Biocosmolog\'{\i}a, Pedro de Alvarado, 14, 06411-Medell\'{\i}n, Spain.}
\date{\today}

\begin{abstract}
We consider a multiverse scenario made up of classically disconnected regions of the space-time that are, nevertheless, in a quantum entangled state. The addition of a scalar field enriches the model and allows us to treat both the inflationary  and the `oscillatory stage' of the universe on the same basis. Imposing suitable boundary conditions on the state of the multiverse, two different representations are constructed related by a Bogoliubov transformation. We compute the thermodynamic magnitudes of the entanglement, such as entropy and energy, explore the effects introduced by the presence of the scalar field and compare with previous results in the absence of scalar field.
\end{abstract}

\maketitle


\section{Introduction}

The search of a satisfactory explanation of the current accelerating stage of the universe has entailed the study of a wide variety of new cosmic scenarios. Among them, the multiverse stands out, probably, as one of the most controversial since it appears to be an untestable proposal. However, this would not be the case if a particular theory provided us with observable and distinguishing predictions of the effects of other universes on the properties of our single universe. That would bring the multiverse into the physical scene of testable theories.  

Different multiverse scenarios can be found in the literature \cite{Carr2007, Tegmark2004, Tegmark2007, Mersini2008, Mersini2008b, Saunders2010}, so we need first specify the kind of multiverse we are dealing with in this paper. We shall consider a multiverse made up of causally disconnected regions of the space-time, each of which will be named throughout the paper with the word `universe'. The universes of the multiverse can be topologically disconnected, i.e., they can be simply-connected regions of a larger multiply-connected manifold, or they can be causally separated by the existence of event horizons that prevent them from any physical signaling. At first sight it would seem that we should just consider one of those regions and disregard the rest of them as physically admissible. However, classical and quantum correlations may appear among different universes of the physical multiverse \cite{PFGD2013} as well as residual interactions coming from the dimensional reduction of multi-dimensional theories \cite{Alonso2012}. In that case, other universes should be considered as well in order to describe physical reality \cite{Mersini2008b}. 

It is worth noticing that the multiverse opens the door to the possibility of having quantum effects with no classical analogue, like entanglement or squeezing \cite{Reid1986, PFGD1992a}, in an otherwise large macroscopic universe. Thus, the quantum effects of the space-time of a single universe may not be only restricted to the very early stage of the universe but they could appear as well on macroscopic scales, becoming therefore testable. 

In the present work, the universes of the multiverse will quantum mechanically be described in the framework of the so-called third quantization formalism \cite{McGuigan1988, Rubakov1988, Strominger1990}, which has recently received a renewed attention \cite{Faizal2012, Faizal2012a, Kim2012, RP2010}. It basically consists of considering the wave function of the universe as the field to be quantized (this field propagates along the variables of the minisuperspace). Then, the general solution of the Wheeler-DeWitt equation can be given in terms of an orthonormal basis of number states that would give, in an appropriate representation, the number of universes of the multiverse.

Such an appropriate representation of universes is, however, a difficult task to elucidate in the multiverse scenario, and it depends on the boundary condition that is imposed on the state of the whole multiverse. Furthermore, the existence of quantum correlations in a composite state crucially depends on the representation chosen \cite{Vedral2006} and it may indeed happen in the multiverse that universes which appear to be independent in a given representation may interact in another representation \cite{Alonso2012}. Therefore, the boundary conditions of the multiverse will eventually determine the correlations in the composite state of different universes.

In this paper, we shall consider two main representations. First, we shall use an invariant representation which is consistent with the general boundary condition that we impose: the  global properties of the multiverse do not depend on the value of the scale factor of a particular single universe. However, an observer inside a single universe would describe her universe in the asymptotic representation of a large parent universe like ours. In this paper, we shall show that these two representations are related by a Bogoliubov transformation and, thus, we can compute the thermodynamic properties of inter-universal entanglement much in a parallel way as they are computed in the context of a quantum field theory in a curved space-time \cite{Hawking1974, Hawking1976, Hawking1976a, Gibbons1977, Lapedes1978}. There, the existence of an event horizon --a black hole horizon in the case of the Schwarzschild metric or a cosmological horizon in the case of a de-Sitter space-time-- makes inaccessible a part of the whole space-time. Then, the observable sector appears to be filled with thermal radiation. In the case of a pair of entangled universes, one of the universes becomes inaccessible to an observer inside the partner universe. Then, it will be shown that, as a result of the inter-universal entanglement, such an observer would perceive her universe as being in a thermal state which is indistinguishable from a classical mixture \cite{Partovi2008} but whose properties depend on the rate of entanglement between the universes. In fact, it actually comes from a sharp quantum state.

The aim of this paper is to study the effects that realistic matter fields may have on the quantum correlations of an entangled pair of universes. These universes can be seen as coming from a double instanton of the Euclidean regime that gives rise to a pair of Lorentzian universes whose global properties are correlated \cite{RP2012b}.

Furthermore, the thermodynamics of entanglement is expected to provide us with a quantum generalization of the customary formulation of thermodynamics \cite{Vedral2002, Brandao2008}, and quantum entanglement may be seen as a novel source of thermodynamic properties \cite{Anders2007, Amico2008}. The development of such an ambitious program would entail a major achievement for the multiverse proposal we are dealing with, considering that it implies that the thermodynamic properties of inter-universal entanglement could eventually  be related to the customary thermodynamic properties of the universe, such as its energy or entropy. If that were the case, we should consider as well inter-universal entanglement in the general thermodynamic picture of the universe. Particularly, it might have significant consequences in the vacuum energy and in the arrow of time of our universe \cite{RP2012b,RP2012}.

The outline of the paper is the following. First we describe in section \ref{section2} the details of the model we are considering and we give the semiclassical solutions of the Wheeler-DeWitt equation for two different scenarios: the slow roll stage of the scalar field, where the potential is approximately constant and contributes to a large value of the cosmological constant, that is, an inflationary stage of the universe; and, on the other hand, we consider the oscillatory regime of the scalar field. In section \ref{section3} we deal with the boundary conditions and with the construction of the two quantum representations relevant for our results. Then, in section \ref{section4} we compute some thermodynamic magnitudes of the entanglement between universes and we discuss the role played by the scalar field and compare with other scenarios without a scalar field. After the conclusions (section \ref{conclusions}) we added an appendix (\ref{app1}) where we show the standard procedure of the quantization of a scalar field in a de-Sitter space-time and the thermal bath derived from it, that has some analogies with the procedure followed in our case.


\section{Semiclassical state of the universe}\label{section2}

We consider a multiverse made up of large homogeneous and isotropic regions of the space-time with closed spatial sections, which are endorsed with a cosmological constant $\Lambda_i$ and a set of $n$ scalar fields, $\vec{\varphi}^{(i)}\equiv (\varphi_1^{(i)},\varphi_2^{(i)}, \ldots, \varphi_n^{(i)})$, that represent the matter content of the $i$-universe. The index $i$ labels the different types of  universes that can exist in the multiverse, which in the present case are all homogeneous and isotropic universes. For instance, in the landscape of the string theories $\Lambda_i$ would run over all the values of the vacua of the landscape. Also notice that homogeneity and isotropy are assumable conditions as far as we work with universes of a length scale well above from the Planck length, where the quantum fluctuations of the space-time can be disregarded. Thus, the homogeneity and isotropic conditions are even valid for most of the Euclidean regime (classically forbidden region) provided that the energy scale for which the universe crosses to the Lorentzian region (classically allowed region) is far from the Planck mass, $M_P\sim 10^{19} {\rm GeV}$. Furthermore, potential observers would presumably inhabit large homogeneous and isotropic regions of the space-time like our single universe that have undergone an inflationary stage, so that anisotropies and inhomogeneities can be disregarded in a first approach in the state of the single universes, although they may play an important role in the global picture of the multiverse \cite{Linde1990}.

For a large homogeneous and isotropic region of the space-time general relativity is effectively valid and the Friedmann-Robertson-Walker
(FRW) metric can locally describe the geometry of the space-time. Then, following the canonical quantization procedure, the single $i$-universe of the multiverse would quantum mechanically be described by a wave function $\phi_i$ defined in the minisuperspace with the set of variables $\{ q^A \}\equiv \{ a, \vec{\varphi}^{(i)} \}$, where $a$ is the scale factor and $\vec{\varphi}^{(i)}\equiv (\varphi_1^{(i)},\varphi_2^{(i)}, \ldots, \varphi_n^{(i)})$ are the $n$-fields that represent the matter content of the $i$-universe. In that case, the wave function $\phi_i\equiv\phi_i(a,\vec{\varphi})$ is the solution of the Wheeler-DeWitt equation
\begin{equation}\label{GeneralWDW}
\left\{ - \nabla_{LB}^2 + \mathcal{V}^{(i)}(a,\vec{\varphi})  \right\} \phi_i(a,\vec{\varphi}) = 0 ,
\end{equation}
where the `Laplace-Beltrami operator' $\nabla_{LB}$ is the covariant generalization of the Laplace operator \cite{Kiefer2007} given by
\begin{equation}\label{LB}
\nabla_{LB}^2 \equiv \frac{1}{\sqrt{-\mathcal{G}}} \frac{\partial}{\partial q^A} \left( \sqrt{-\mathcal{G}} \, \mathcal{G}^{AB} \frac{\partial}{\partial q^B}  \right),
\end{equation}
and $\mathcal{V}^{(i)}(a,\vec{\varphi})$ is the potential of each of the fields. The minisupermetric $\mathcal{G}_{AB}$ possesses a Lorentzian signature \cite{Kiefer2007}, namely  $\mathcal{G}_{AB} \equiv {\rm diag}(- a, a^3, \ldots,  a^3)$ in appropriate units. It allows us to set down a formal analogy between the Wheeler-DeWitt equation (\ref{GeneralWDW}) and the wave equation for a field that propagates in a curved space-time.  The scale factor, $a$,  formally plays the role of an intrinsic time variable and the matter fields, $\vec{\varphi}^{(i)}$ of the $i$-universe, the role of the spatial components. Then, we can study the quantum state of the multiverse in the framework of a quantum field theory in the $n+1$-dimensional minisuperspace determined by the minisupermetric $\mathcal{G}_{AB}$.

The general quantum state of the multiverse is given by a wave function, $\Psi_{\vec{N}}(a, \vec{\phi})$, which is a linear combination of product states like \cite{RP2012b}
\begin{equation}\label{stateMultiverse}
\Psi_{N_1}^{\vec{\alpha}_1}(a, \phi_1)  \Psi_{N_2}^{\vec{\alpha}_2}(a, \phi_2) \cdots  \Psi_{N_m}^{\vec{\alpha}_m}(a, \phi_m) ,
\end{equation}
where $\vec{\phi}\equiv (\phi_1, \phi_2, \ldots, \phi_m)$, and $\vec{N}\equiv (N_1, N_2, \ldots, N_m)$, with $N_i$ being the number of universes of type $i$ represented by the wave function $\phi_i \equiv \phi(a, \vec{\varphi}_i)$ that corresponds to a universe which is described in terms of $\vec{\varphi}_i$ matter fields and $\vec{\alpha}_i\equiv (\alpha_{i,1}, \ldots, \alpha_{i,k})$ parameters. In the context of the landscape, for instance, the functions $\Psi_{N_i}^{\Lambda_i}(a, \phi_i)$ in Eq. (\ref{stateMultiverse}) would be the solutions of the third quantized Schr\"{o}dinger equation \cite{RP2012b}
\begin{equation}\label{3Schrodinger}
i \frac{\partial}{\partial a} \Psi_{N_i}^{\Lambda_i}(a, \phi_i) = \cH^{(i)}(a,\phi, p_\phi) \Psi_{N_i}^{\Lambda_i}(a, \phi_i) ,
\end{equation}
where $\cH^{(i)}(a,\phi, p_\phi)$ is the third quantized Hamiltonian \cite{Strominger1990, RP2010} that corresponds to each kind of universe, with $p_\phi \equiv \sqrt{-G} G^{0 B} \nabla_B\phi$\, and $\nabla_B$ being the third quantized momentum and the covariant derivative in the minisuperspace respectively. Throughout this paper we shall work in units for which $\hbar = 1 = c$. Let us notice that we could consider as well Hamiltonians of interaction between different species of universes adding a more exhaustive phenomenology to the model of the multiverse \cite{Alonso2012}. 

In the present model, considering for simplicity only one scalar field, the Wheeler-DeWitt equation (\ref{GeneralWDW}) for the $i$-universe can be written as \cite{Linde1990, Kiefer2007}
\begin{equation}\label{GeneralWDW1}
 \ddot{\phi} +  \frac{\dot{\mathcal{M}}(a)}{\mathcal{M}(a)} \dot{\phi} - \frac{1}{a^2} \phi'' + \omega^2(a,\varphi) \phi = 0 ,
\end{equation}
where the scalar field has been rescaled as $\varphi \rightarrow \f{2}{M_P}\sqrt{\f{\pi}{3}}\varphi$, with $M_P$ the Planck mass, $\dot{\phi}\equiv \f{\pp\phi}{\pp a}$ and $\phi' \equiv \f{\pp\phi}{\pp\vphi}$, with $\phi\equiv\phi(a,\vphi)$ and $\cM(a)\equiv a$, and therefore $\dot{\cM}(a)=1$ (we choose this notation in order to ease the analogy, afterwards, with the harmonic oscillator). In the units we are working with, $\varphi$ has units of mass and it thus turns out to be dimensionless after rescaling. For clarity, the index $i$ of the $i$-universe has been removed assuming that all the expressions throughout this section will be given for a single universe unless otherwise indicated. The potential term in Eq. (\ref{GeneralWDW1}) can generally be written as,
\begin{equation}\label{GeneralFrequency}
\omega^2(a,\vphi) \equiv \sigma^2 (H^2 a^4 - a^2 ) ,
\end{equation}
where $\sigma \equiv \f{3 \pi M_P^2}{2}$ and $H\equiv H(\varphi)$ is the Hubble function. The frequency $\omega$ has units of mass or, equivalently, units of the inverse of time or length, as it was expected. We shall consider two contributions to the Hubble function, i.e., $H^2 = H_0^2 + H_1^2$. The first one is caused by the existence of a cosmological constant, $\Lambda_0 = 3 H_0^2$, which is assumed to be very small. The second contribution is due to the potential of the scalar field, $H_1^2 = \frac{8 \pi}{3 M_P^2} V(\varphi) = m^2 \varphi^2$, where in the last equality it has been assumed a quadratic potential.

As it is well-known \cite{Linde1990, Linde2007}, there are four scales of interest in the creation and subsequent inflationary picture of a single universe (in the following we shall assume a quadratic potential although the picture is rather general, see Refs. \cite{Linde1990, Linde2007}). First, for values $\varphi \gtrsim \lambda^{-1}$ (i.e., $V(\varphi) \gtrsim M_P^4$), with $\lambda \equiv \frac{m}{M_P}\ll 1$, there is no consistent description of the space-time because the quantum fluctuations of the metric tensor become of the same order than the components of the metric. It corresponds to the realm of the space-time foam \cite{Wheeler1957, Hawking1978, Garay1998}. For the value $\lambda^{-1} \gtrsim \varphi \gtrsim \lambda^{-\f12}$ ($M^4_P \gtrsim V(\varphi) \gtrsim \lambda M_P^4$), the fluctuations of the space-time are weakened and the large value of the scalar field makes it to slowly roll down the potential during the time scale $H^{-1} \gtrsim M_P^{-1}$. The energy of the field is stored in the potential term, which is approximately a constant, and inflation then starts in the Lorentzian regions of the space-time. Besides, the fluctuations of the scalar field are large and give rise to new inflationary regions, i.e., new universes are nucleating in an eternal `self-inflationary' process \cite{Linde1983, Linde1986}.

For a value $\lambda^{-\f12} \gtrsim \varphi \gtrsim 1$ ($\lambda M_P^4 \gtrsim V(\varphi) \gtrsim \lambda^2 M_P^4$), inflation goes on but the fluctuations of the scalar field are weakened and the creation of new universes stops. Finally, for values $1 \gtrsim \varphi \gtrsim 0$ ($ \lambda^2 M_P^4 \gtrsim V(\varphi) \gtrsim 0$), the slow-roll approximation fails and, after a `graceful exit' \cite{Mukhanov2005}, the scalar field decays into the particles that will conform the actual structure of the universe. Most of the energy stored in the scalar field during the inflationary stage is now transmitted to the particles and the universe enters thus in the hot regime \cite{Linde1990, Mukhanov2005, Linde2007}.

Our model will quantum mechanically picture the following scenario. First, we shall consider the quantum creation of universes in entangled pairs. After being created, these universes undergo an inflationary stage in which their quantum states are still correlated. The exponential expansion of the universe is then led by an effective value $H_1^2$ of the cosmological constant, because $\Lambda_0 \ll1$. After the exit of inflation, the universe enters in the oscillatory regime of a scalar field which is quantized in the curved background of a de-Sitter space-time with a value $\Lambda_0$ of the cosmological constant. The wave function of the universes may still retain some residual correlations that might be tested in the current stage of the universe.

\begin{figure}
\centering
\includegraphics[width=8cm]{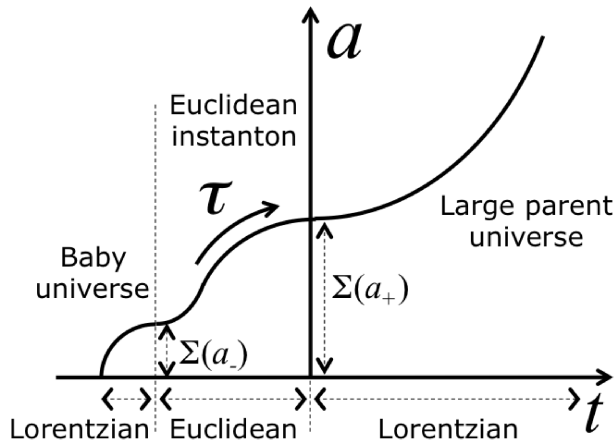}
\caption{The creation of a large parent universe from a baby universe.}
\label{fig1}
\end{figure}

\subsection{Creation of universes in entangled pairs}

Within a region slightly larger than the Planck scale the space-time can approximately be considered homogeneous and isotropic provided that the energy associated to the scalar field is smaller than the Planck energy density, i.e., $V(\varphi) \lesssim M_P^4$. Then, for a sufficiently uniform and slowly varying field with a large initial value ($\varphi_0 \gg 1$), the equations of motion for the scale factor and the scalar field read \cite{Linde1990}
\be\label{eqda}
\f{d a}{d t} = \f{\omega}{\sigma a} \approx \sqrt{a^2 H_1^2 - 1} \,,\,\,\qquad \,\, \f{d \varphi}{d t} \approx -\f{1}{3 H_1} \f{d V}{d \varphi} .
\ee
For a value $a < H_1^{-1}$ there is no Lorentzian solution to the first of these equations. However, by performing a Wick rotation to Euclidean time $\tau$, the Euclidean solution
\be
a_E(\tau) = \f{1}{H_1} \cos (H_1\tau) ,
\ee
with $\tau \in (-\f{\pi}{2 H_1}, 0)$, represents the scale factor of a de-Sitter instanton that shrinks to zero as the Euclidean time approaches the value $\tau \rightarrow -\f{\pi}{2 H_1}$. Before reaching the singular value $a_E = 0$ the Euclidean instanton would delve into the space-time foam where the approximations of homogeneity and isotropy are, actually, no longer valid. At the Euclidean time $\tau = 0$ the instanton finds the Lorentzian regime and the universe locally emerges as a Friedmann-Robertson-Walker space-time with a scale factor given by
\be
a(t) = \f{1}{H_1} \cosh (H_1 t) ,
\ee
with $t\geq 0$.

Quantum mechanically, the wave function of the universe can be decomposed in normal modes \cite{RP2010, RP2012b},
\be\label{modedecomposition}
\phi(a,\varphi) = \int dk \; e^{i k \varphi} \phi_k(a) \, \hat{c}_{0,k} + e^{-i k \varphi} \phi_k^*(a) \, \hat{c}_{0,k}^\dag ,
\ee
where $\hat{c}_{0,k}^\dag$ and $\hat{c}_{0,k}$ are constant operators that represent, respectively, the creation and annihilation of universes with a particular value $k$ of the mode. Inserting the wave function (\ref{modedecomposition}) into the Wheeler-deWitt equation (\ref{GeneralWDW1}), it follows that the probability amplitudes $ \phi_k(a)$ satisfy the equation of the damped harmonic oscillator
\be\label{ho01}
\ddot{\phi}_k + \frac{\dot{\cM}}{\cM} \dot{\phi}_k + \omega_k^2 \phi_k = 0 ,
\ee
with a scale factor dependent frequency given by
\be\label{fre}
\omega_k\equiv \omega_k(a) = \sigma \sqrt{a^4 H_{1}^2 - a^2 + \frac{k^2}{\sigma^2 a^2}}  = \frac{\sigma H_{1}}{a} \sqrt{(a^2 - a_+^2)(a^2 - a_-^2)(a^2 + a_0^2)} ,
\ee
where 
\beq
a_+ \equiv a_+(k) &=& \frac{1}{\sqrt{3}H_{1}} \sqrt{1 + 2 \cos(\frac{\alpha_k}{3})} , \\
a_- \equiv a_-(k) &=& \frac{1}{\sqrt{3 } H_{1}} \sqrt{1 - 2 \cos(\frac{\alpha_k + \pi}{3})} , \\
a_0 \,\equiv\, a_0(k) &=& \frac{1}{\sqrt{3}H_{1}} \sqrt{-1 + 2 \cos(\frac{\alpha_k - \pi}{3})} ,
\eeq
with $a_+ \geq a_- \geq a_0$ and
\be
\alpha_k \equiv {\rm arccos}(1-\frac{2 k^2}{k_m^2}) \in [0,\pi] ,
\ee
considering $k_m^2 \equiv \frac{4 \sigma^2}{27 H_{1}^4} = \f{\pi^2 M_P^4}{3 H_1^4}$ and $k_m \geq k \geq 0$. It is worth noticing that the value $k$ of the mode is related to the momentum $p_\varphi$ that is classically proportional to $\partial_t\varphi$. In the slow-roll approximation we should classically consider the value $k= 0$ that corresponds to the ground state. However, as we have already pointed out, the fluctuations of the scalar field can be very large during the first stage of the inflationary period, when universes are created, so we have to quantum mechanically consider as well other modes for the wave functions of the universes being created.

\begin{figure}
\centering
\includegraphics[width=8cm]{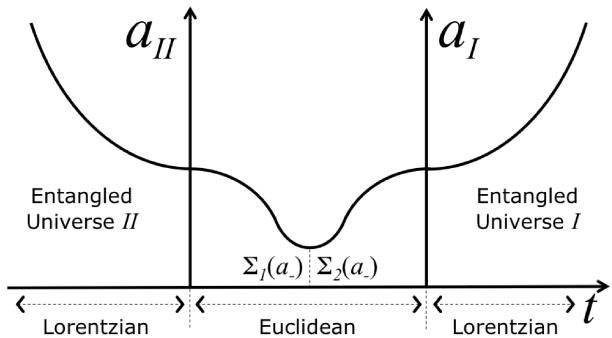}
\caption{The creation of a pair of entangled universes from a double instanton.}
\label{fig2}
\end{figure}

The consideration of other modes different from zero entails a significant departure from the customary picture of the creation of universes because of the quantum correction that appears in the frequency (\ref{fre}). Unlike in Eq. (\ref{eqda}), where there is one Euclidean region and one Lorentzian region,  in the equation of motion corresponding to the frequency given by Eq. (\ref{fre}) there is one Euclidean region, for the value $a_+ > a > a_-$, between two Lorentzian regions, for values $a>a_+$ and $a<a_-$. Therefore, for $k_m>k>0$ there are two possibilities for the birth of the universe \cite{RP2012b}, which are represented in Figs. \ref{fig1}-\ref{fig2}. In Fig. \ref{fig1} it is represented the birth of the universe from a preexisting baby universe. It also exists the possibility that a pair of entangled universes is created from the double instanton that is formed by joining two single instantons (like it is proposed in Ref. \cite{Barvinsky2006}), as it is depicted in Fig. \ref{fig2}. In this case, the two universes must be in an entangled state due to the fact that the Euclidean instantons can only be matched for the same value of  $|k|$ so, therefore, the modes of the wave function of the corresponding universes are correlated.

Let us also notice that, like in the customary case of a universe being created from \emph{nothing} \cite{Vilenkin1982, Vilenkin1984}, the approximations of homogeneity and isotropy are still valid in the case of the creation of universes in entangled pairs provided that $a_-$ is  large enough with respect to the Planck length. Considering a plausible value of the mass of the scalar field of $m \sim 10^{15} {\rm GeV}$ (i.e., the GUT scale), then $\lambda \sim 10^{-4}$ and the values of $H_1$ for which the universes are created are $M_P^2 > H_1^2 \gtrsim \lambda M_P^2$. It implies that the size of the newborn universes is of order $\ell_{H_1} \sim H_1^{-1}$, with $\ell_{H_1} \sim \lambda^{-\f12} \ell_P \sim 10^2 \ell_P$. Thus, there is still room for the creation of a pair of homogeneous and isotropic universes with, $10^2 \ell_P \approx a_+ > a_- \gg \ell_P$; the range of possible values could be even larger for smaller values of $\lambda$.

\subsection{Inflationary stage of the universes}

Once the universes have entered into the inflationary regime their scale factors exponentially grow by an overall factor \cite{Linde1990} $P\sim e^{\lambda^{-2}}$ until the slow-roll approximation eventually fails at $\varphi \sim 1$. After the initial stages of inflation the quantum correction term in Eq. (\ref{fre}) is quantitatively negligible. The solutions of the Wheeler-deWitt equation (\ref{ho01}) can then be written in the WKB approximation as
\begin{equation}
\phi_k(a) = \frac{1}{\sqrt{\cM(a) \, \omega_k(a)}} e^{\pm i \sigma \mathcal{S}_k(a)} \approx \frac{1}{\sqrt{\cM(a) \, \omega(a)}} e^{\pm i \sigma \mathcal{S}(a)} ,
\end{equation}
where 
\begin{equation}\label{Sk}
\mathcal{S}_k(a) \equiv \f{1}{\sigma} \int^a da' \, \omega_k(a') \approx \f{1}{\sigma} \int^a da' \, \omega(a') \equiv S(a)  ,
\end{equation}
with $\omega(a)$ given by Eq. (\ref{GeneralFrequency}). 

Although quantitatively negligible, it is worth pointing out that the quantum correction term of Eq. (\ref{fre}) for the $k$ mode is the ultimate reason for the plausible creation of universes in entangled pairs and therefore for the correlations between universes. The classical evolution of each single universe is given however by a Friedman equation, $\partial_t a = \f{\omega(a)}{\sigma a} \approx  H_1 a$, which is effectively independent of the value of the mode $k$, i.e., all the universes corresponding to different modes classically evolve in a similar way irrespective of the quantum mode from which they were created. They undergo an exponential expansion given by $a(t) \approx a_0 \, e^{H_1 t}$.

While the universes are expanding, the field is slow-rolling down to the minimum of the potential until the scalar field approaches the value $\varphi \sim 1$. Then, the slow-roll approximation fails, the Hubble parameter $H_1$ is no longer large enough to prevent the field from rapidly rolling down to the minimum of the effective potential where it starts to oscillate. The energy of the field, so far stored in the potential term $V(\varphi)$, is now transferred to the particles that are created as a result of the oscillations of the field. These particles eventually collide between them and the universe heats up \cite{Linde1990}. The hot universe cools down and the initial particles decay into new particles that will conform the structure of the current universe.

\subsection{Oscillatory regime of the scalar field}

Each single universe of the entangled pair (in the scenario represented in Fig. \ref{fig2}) leaves the inflationary stage when the slow-roll approximation fails. The scalar field enters then in the oscillatory regime and it becomes meaningful to describe it in terms of particles. In the oscillatory regime, the semiclassical solutions of the Wheeler-DeWitt equation for each single universe, Eq. (\ref{GeneralWDW1}) with the frequency of the Eq. (\ref{GeneralFrequency}), can be written as \cite{Hartle1990, Hartle1993}
\begin{equation}
\label{GeneralSC}
\phi^{sc}(a,\vphi) = \frac{1}{\sqrt{\cM(a) \, \omega_0(a)}} e^{\pm i \sigma \mathcal{S}(a)} \Delta(a,\vphi) ,
\end{equation}
with a prefactor that depends only on the gravitational degrees of freedom (only on the scale factor), where $\omega_0$ is given by Eq. (\ref{GeneralFrequency}) with $H = H_0$, 
\begin{equation}
\label{A1}
\mathcal{S}(a) \equiv \f{1}{\sigma} \int^a da' \, \omega_0(a') = \f{ (a^2 H_0^2 - 1)^\f{3}{2}}{3 H_0^2} ,
\end{equation}
and the function $\Delta(a,\vphi)$ contains all the information about the scalar field. In the semiclassical regime, this function  satisfies the Schr\"{o}dinger equation \cite{Hartle1990}
\begin{equation}
\label{Schrodinger}
\mp i \f{\omega_0(a)}{\sigma a} \dot{\Delta}(a,\vphi) = \left( -\f{1}{2 \sigma a^3} \f{\pp^2}{\pp\vphi^2} + 2 \pi^2 a^3 V(\varphi)\right)\Delta(a,\vphi),
\end{equation}
where $\dot{\Delta}\equiv\f{\pp\Delta}{\pp a}$.  The expansion (or contraction) of the universe, given by the Friedmann equation 
$\f{\pp a}{\pp t} = \mp \f{\omega_0(a)}{\sigma a}$, where 
$t$ is the Friedmann time, provides us with a time variable, $t$, that  is well-defined in each branch of the universe \cite{Vilenkin1986}. In terms of the Friedmann time, $t$, the Schr\"{o}dinger equation (\ref{Schrodinger}) takes the customary form
\begin{equation}\label{Schrodinger2}
i  \frac{\partial \Delta}{\partial t}(\varphi, t) = \hh(\varphi, t) \Delta(\varphi, t) ,
\end{equation}
where $\hh(\varphi,t)$ is the corresponding Hamiltonian for the matter field $\varphi$. We can observe that this equation turns out to be the  Schr\"{o}dinger equation for a scalar field in the semiclassical background of a de-Sitter space-time with a value $\Lambda_0\ll1$ of the cosmological constant (let us recall that $H^2_0 + H_1^2 \gg1$ only because $H_1^2 \gg1$ during inflation).

It is worth noticing that the equations (\ref{GeneralSC}-\ref{Schrodinger2}) are rather general and they are valid for other kind of potentials and matter fields  of the $i$-type of universes in the multiverse (the generalization of Eq. (\ref{GeneralSC}) to spinorial fields is straightforward). Therefore, the effects of other types of fields can be studied following the same general procedure used in this paper. The most general quantum state of the multiverse would be given by a linear combination  of product states (\ref{stateMultiverse}) relating all types of universes of the multiverse with different kind of fields that represent the matter-energy content of the $i$-universe. For the sake of concreteness, let us particularize the general semiclassical solution (\ref{GeneralSC}) to a scalar field with mass $m$ and $V(\varphi) = \f12 m^2 \varphi^2$ (notice that\, $V(\varphi)= \f{\sigma}{4 \pi^2} m^2 \varphi^2$ after rescaling the field), with $H_1^2 \gg m^2 \gg H_0^2$. Then, $\hh$ in Eq. (\ref{Schrodinger2}) reads
\begin{equation}\label{hamiltonianh}
\hh \equiv \frac{1}{2 M(t)} p_\vphi^2 + \f{M(t) m^2}{2} \varphi^2 ,
\end{equation}
where $M(t) \equiv \sigma a^3(t)$ and $p_\varphi\equiv - i \f\pp{\pp\vphi}$. Notice that the Hamiltonian (\ref{hamiltonianh}) leads to the customary quantization of a scalar field with mass in the curved background of the de-Sitter space-time (see Appendix \ref{app1}).

The function $\Delta(t,\vphi)$ in Eq. (\ref{Schrodinger2}) can be expressed in terms of the eigenfunctions $\Delta_n(t,\vphi)$ of the harmonic oscillator with time dependent mass \cite{Lewis1969, Dekker1981, Jannussis1988, Dantas1992, Kim2001, Vergel2009}, i.e., $\Delta(t,\vphi) =\sum_n B_n \Delta_n(t,\vphi)$, with $B_n$ constant coefficients and $\Delta_n(t,\vphi)$ the normalized eigenfunctions that  can be expressed as (see, for instance, Sec. 4.2 of Ref. \cite{Dekker1981}) 
\begin{widetext}
\begin{equation}
\Delta_n(t,\vphi)=\left(\f{1}{2^n n!}\sqrt{\f{M(t)\,\widetilde m}{\pi}} \right)^\f12 e^{-i(n+\f12)\widetilde m t}e^{\f{i M(t)}{2}(-\f{3 H_0}{2}+i \widetilde m) \vphi^2}  {\rm H}_n(\sqrt{M(t)\,\widetilde m}\,\vphi),
\end{equation}
\end{widetext}
with $M(t) \approx \frac{\sigma}{8 H_0^3} e^{3 H_0 t}$, ${\rm H}_n(x)$ the Hermite polynomial of degree $n$, and $\widetilde m = \sqrt{m^2 - \frac{9 H_0^2}{4}}$, ($\widetilde m \approx m $ for the subdamped regime of the harmonic oscillator, i.e., for values $m \gg H_0$). In the previous equation, we have made use of the approximation $a\gg H_0^{-1}$, that is valid in the semiclassical approximation that we are considering. With such an approximation, the value of the scale factor coming from the Friedmann equation $\f{\pp a}{\pp t} = \mp\f{\omega_0(a)}{\sigma a}$ takes the form: $a(t) = H_0^{-1} \cosh H_0 t\approx \f{1}{2 H_0} e^{H_0 t}$. Then, by applying the inverse relation $t  \approx H_0^{-1} \ln(2H_0a)$, we can finally write the semiclassical solutions of the wave function of the universe in terms of the scale factor and the scalar field as
\begin{equation}
\phi^{sc}(a,\vphi) = \sum_n B_n \phi_n^{sc}(a,\varphi) \equiv \sum_n B_n |\phi^{sc}_n(a,\vphi)| e^{\pm i \sigma S_n(a,\vphi)} ,
\end{equation}
where
$S_n(a,\vphi) \approx \f{ H_0 a^3}{3} (1 + \f{9}{4} \vphi^2) + \f{\widetilde m}{\sigma H_0} (n+\frac{1}{2}) \ln(H_0 a)$,
and
\begin{widetext}
\begin{equation}
|\phi^{sc}_n(a,\vphi)|\approx\f{1}{\sqrt{2^n n!}}\left(\f{\widetilde m}{\pi \sigma H_0^2 a^3} \right)^\f{1}{4}e^{-\f{ \sigma \widetilde m }{2} a^3 \vphi^2} {\rm H}_n(\sqrt{\sigma \widetilde m} \, a^\f{3}{2} \vphi).
\end{equation} 
\end{widetext}

Furthermore, with the help of Eq. (\ref{GeneralSC}),  the Wheeler-DeWitt equation (\ref{GeneralWDW1}) can be rewritten for each mode, in the semiclassical regime as
\begin{equation}\label{GeneralWDW3}
 \ddot{\phi}_n +  \frac{\dot{\mathcal{M}}(a)}{\mathcal{M}(a)} \dot{\phi}_n + \Omega^2_n(a,\varphi) \phi_n = 0 ,
\end{equation}
with
\begin{equation}
\Omega^2_n(a,\varphi) = \omega_0^2 + i \frac{2 \sigma a}{\Delta_n} \frac{\partial \Delta_n(t,\varphi)}{\partial t}\Big|_{t=t(a)}=\omega_0^2\mp 2i\omega_0\frac{\dot{\Delta}_n}{\Delta_n} .
\end{equation}
The $\mp$ sign will be chosen depending on whether we consider an expanding or a contracting branch of the universe. In order to study the properties of the multiverse in the semiclassical approximation, it is useful to write down the leading order of the asymptotic behavior of $\Omega_n(a)$ for large values of the scale factor $a$:
\be
\label{asintoticOmega}
\Omega_n \sim \Omega \equiv \left(H_0^2 (1\mp\frac{9}{2}\varphi^2)\pm i3H_0\widetilde m\varphi^2\right)^{1/2} \sigma a^2.
\ee

In the third quantization formalism, Eq. (\ref{GeneralWDW3}) can  formally be seen as the equation of an harmonic oscillator placed in a dispersive medium with \emph{time} dependent mass and frequency, $\cM\equiv\cM(a)$ and $\Omega_n \equiv\Omega_n(a,\vphi)$, where the scale factor  plays the role of the intrinsic time variable and the scalar field that  of a spatial variable. Nevertheless, the frequency $\Omega_n=|\Omega_n|e^{i\theta_n}$ is, in general, complex. Using Eq. (\ref{asintoticOmega}), we obtain the polar angle $\theta_n$ does not depend on $a$ at the leading order: $\dot{\theta}_n\approx \dot{\theta} =0$. 

We would like to have a pure real frequency in order to ease the analogy with a proper harmonic oscillator and define, afterwards, the associated creation and annihilation operators. The WKB solution to the Wheeler-DeWitt equation takes the form:
\begin{equation}
\phi^{sc}_n(a,\vphi)=\f{1}{\sqrt{\cM(a)\Omega_n(a,\vphi)}}e^{\pm i S^{(n)}(a,\vphi)}\,,
\end{equation}
with $\dot{S}^{(n)}(a,\vphi) \equiv \Omega_n$. This expression can be written in the following equivalent way:
\begin{equation}
\phi^{sc}_n(a,\vphi)=\f{1}{\sqrt{\cN_n(a)\omega_n(a,\vphi)}}e^{\pm i S_{\omega}^{(n)}(a,\vphi)}\,,\label{solutionsc}
\end{equation}
with  $\dot{S}_{\omega}^{(n)}(a,\vphi) \equiv \omega_n$, and
$$
\omega_n =|\Omega_n|\cos \theta_n\,,\quad
\cN_n(a)=\f{\cM(a)}{|\cos\theta_n|}e^{\pm 2 S_{I}^{(n)}},
$$
where $S_I^{(n)}\equiv \Im (S^{(n)})=\int |\Omega_n|\sin \theta_n da$, is the imaginary part of the action $S^{(n)}$. Now both $\omega_n \in \R$ and $\cN_n(a) \in \R$. Besides, the solution (\ref{solutionsc}) satisfies the following dumped harmonic oscillator equation:
\begin{equation}\label{GeneralWDW4}
\ddot{\phi}^{sc}_n +  \f{\dot{\cN}_n(a)}{\cN_n(a)} \dot{\phi}^{sc}_n + \omega_n^2(a,\vphi) \phi^{sc}_n = 0 ,
\end{equation}
that can be derived from the Hamiltonian
\begin{equation}\label{Hamiltonian}
\cH_n = \f{1}{2 \cN_n(a,\vphi)} p_{\phi_n}^2 + \f{\cN_n(a,\vphi) \omega_n^2}{2} \phi_n^2.
\end{equation}
The total Hamiltonian of the $i$-type of universes, $\cH^{(i)}$ in Eq. (\ref{3Schrodinger}), corresponds to the sum of all the contributions of the modes, i.e., $\cH^{(i)} \equiv \sum_n \cH_n^{(i)}$, where the label $i$ of the type of universe has now been reintroduced for explicitness. In the following sections, we will concentrate in one of the modes (of just one type of universe) and the Hamiltonian (\ref{Hamiltonian}) will be used to determine the evolution of the semiclassical state of a single universe.


\section{Boundary conditions and invariant states of the multiverse}\label{section3}

For definiteness, we shall consider from now on just one single $n$-mode of the $i$-type of universes in the multiverse. However, as it has been pointed out previously, the same procedure that we will use in this section can be applied to other modes of the $i$-universe as well as to the rest of types of universes of the multiverse. It is worth noticing that the multiverse turns out  to be thus an enormously rich scenario interpretable in terms of a `particle-soup' of universes, whose complete phenomenology is still to be uncovered.

Following the new point of view offered at the end of the previous section we shall consider that our $n$-mode of the $i$-type of universe (the index $i$ will be omitted throughout the rest of the section) is described by an harmonic oscillator whose mass $\cN_n$ and frequency $\omega_n$ depend on the scale factor $a$ which is, as it has been commented previously, the time-like coordinate of the minisuperspace. The equation of motion and the Hamiltonian of the system are given by Eqs. (\ref{GeneralWDW4}-\ref{Hamiltonian}), respectively. On the basis of the `third quantization' formalism, the quantum state of each universe of the multiverse will be given in terms of the states of an harmonic oscillator, for any kind of potential of the field $V(\vphi)$ (the potential $V(\vphi)$ is related to the frequency $\omega$).

There are two representations of the Hamiltonian (\ref{Hamiltonian}) that can naturally be chosen with a sensible physical interpretation. On the one hand we will consider the usual creation and annihilation operators  of the harmonic oscillator $c_n(a), c_n\dagg(a)$ that diagonalize the Hamiltonian. On the other hand, we construct other kind of creation and annihilation operators $b_n(a), b_n\dagg(a)$ in such a way that the number operator constructed with them is constant under the change of the scale factor. Notice that in the case where the frequency and mass of the harmonic oscillator do not depend on the time, these two representations coincide. Nevertheless, that is not our case.

Let us start with the ``diagonal representation'', i.e., the representation that diagonalizes the Hamiltonian at a given value $a_0$ of the scale factor for each value $n$ of the normal mode, $\hat{c}_{0,n}\equiv \hat{c}_n(a_0)$ and $\hat{c}^\dag_{0,n}\equiv \hat{c}_n^\dag(a_0)$,  with
\begin{eqnarray}
\label{c1}
\hat{c}_n &=& \sqrt{\f{\cN_n \omega_n}{2 }} \left(\hat{\phi}_n + \f{i}{\cN_n \omega_n} \hat{p}_{\phi_n}\right) , \\ \label{c2}
\hat{c}_n^\dag&=&\sqrt{\f{\cN_n \omega_n}{2}}\left(\hat{\phi}_n-\f{i}{\cN_n \omega_n} \hat{p}_{\phi_n}\right).
\end{eqnarray}
In the third quantization formalism, the wave function of the universe, $\phi(a,\varphi)$, is promoted to an operator that can be written as
\begin{equation}\label{decomposition}
\hat{\phi}(a,\vphi)=\sum_n A_n(a, \vphi) \hat{c}_{0,n} + A_n^*(a,\vphi)\hat{c}^\dag_{0,n} ,
\end{equation}
where the probability amplitudes, $A_n(a, \varphi)$ and $A_n^*(a,\vphi)$ in Eq. (\ref{decomposition}), satisfy the Wheeler-DeWitt equation (\ref{GeneralWDW3}), or equivalently Eq. (\ref{GeneralWDW4}).

However, the operators $\hat{c}_{0,n}^\dag(a)$ and $\hat{c}_{0,n}(a)$ in Eq. (\ref{decomposition}) cannot properly be interpreted as the creation and annihilation operators of universes because, in such representation, the number operator given by $\hat{N}_{c_{0,n}}\equiv \hat{c}_{0,n}^\dag(a) \hat{c}_{0,n}(a)$, is not an invariant operator, i.e.,
$\f{d \hat{N}_{c_{0,n}}}{d a}= i  [\hat{\cH}_n , \hat{N}_{c_{0,n}}]  \neq 0$, 
where $\hat{\cH}_n$, given by Eq. (\ref{Hamiltonian}), is the third quantized Hamiltonian that determines the evolution of the state of the $n$-mode. It means that the excitation number of universes within the multiverse, which is given by the eigenvalues of the number operator, would depend on the value of the scale factor in a particular single universe. This is not expected in the quantum multiverse and a different representation has to be chosen in order to have a parallel notion to that of the creation and annihilation of particles. Notice, that we might have considered the creation and annihilation operators of universal states with the proper frequency of the Hamiltonian, given directly by Eqs. (\ref{c1}-\ref{c2}) for any value of the scale factor, for which $\hat{\cH}_n= \omega_n(a) (\hat{N}_{c,n} +\f12)$ and $[\hat{\cH}_{c,n},\hat{N}_{c,n}]=0$. Nevertheless, the number of universes would be scale factor dependent too, because in this case $\pp_a\hat{N}_c\neq 0$.

An invariant representation, $\hat{b}_n(a)$ and $\hat{b}_n^\dag(a)$, is constructed such that the eigenvalues of the associated number operator, $\hat{N}_{b,n}(a)\equiv \hat{b}_n^\dagger(a)\hat{b}_n(a)$, are invariant under the evolution of the scale factor $a$. That is,
$$
\f{d}{da}\hat{N}_{b,n}(a)=\f{\pp}{\pp a}\hat{N}_{b,n}(a)+ i[\hat{\cH}_n(a), \hat{N}_{b,n}(a)]=0.
$$
The boundary condition we are imposing on the state of the multiverse (the global properties of the multiverse must not depend on the value of the scale factor) fixes the representation that has to be chosen. We look for a representation for which the number operator is invariant. This can be given by the so called Lewis representation \cite{Lewis1969} (see also, Refs. \cite{Pedrosa1987, Dantas1992, Sheng1995, Kim2001}), for which the creation and annihilation operators are defined for each single mode $n$ as \cite{RP2012b}
\begin{eqnarray}\label{b1}
\hat{b}_n(a) &\equiv& \sqrt{\f{1}{2 }} \left( \f{\hat{\phi}_n}{R} + i (R \hat{p}_{\phi_n} - \cN_n\dot{R}\hat{\phi}_n)\right) , \\ 
\label{b2}
\hat{b}_n^\dag(a) &\equiv& \sqrt{\f{1}{2 }} \left( \f{\hat{\phi}_n}{R} - i (R \hat{p}_{\phi_n} - \cN_n\dot{R}\hat{\phi}_n) \right) ,
\end{eqnarray}
with $R \equiv R_n(a)= \sqrt{\phi_1^2(a)+\phi_2^2(a)}$, where $\phi_1$ and $\phi_2$ are two linearly independent solutions of the Wheeler-DeWitt equation (\ref{GeneralWDW4}) that make real the value of the function $R(a)$\footnote{This is to ensure the Hermitian condition of the invariant operator $\hat{I}_n\equiv \hat{N}_n +\f{1}{2}$.},
($R(a)\in\R$). The Lewis representation given by the operators (\ref{b1}-\ref{b2}) conserves the excitation number, i.e., $\hat{b}_n^\dag \hat{b}_n |N,a\rangle = N_n |N,a\rangle$, with $N_n\neq N_n(a)$. Then, the operators $\hat{b}^\dag_n$ and $\hat{b}_n$ can properly be interpreted as the ladder operators of the $n$-mode of the $i$-type of universes in the multiverse. Let us notice that the mode is expected to remain at the value $n$ because we are not  considering interactions between different modes of the wave function of the $i$-universe, i.e., there are no interaction terms in the Hamiltonian of the $i$-universe, $\cH^{(i)} = \sum_n \cH_n^{(i)}$.

However, in terms of the creation and annihilation operators of the Lewis representation, the Hamiltonian turns out to be 
\begin{equation}\label{H2}
\hat{\cH}(a) = \sum_n \hat{\cH}_n =   \sum_n\left( \beta_- \hat{b}_n \hat{b}_{-n} + \beta_+ \hat{b}_n^\dag \hat{b}^\dag_{-n} + \beta_0(a) \left(\hat{b}_n^\dag(a) \hat{b}_n(a) + \f12 \right) \right) ,
\end{equation}
where, $\beta_\pm$ and $\beta_0$ are two non-trivial functions of $R$ \cite{RP2010, RP2012b}. The structure of the Hamiltonian (\ref{H2}) is formally the same to that used in quantum optics to represent the creation and annihilation of entangled pairs of photons \cite{Yuen1976, Reid1986, Scully1997, Walls2008}. This analogy, together with the isotropy of the minisuperspace that suggests that the universes are created in pairs with opposite momenta $n$ and $-n$, allows us to interpret the multiverse as made up of entangled pairs of universes whose properties of entanglement can be analyzed. Notice also that for the case $n=0$ we still have two different entangled universes ($\hat{b}_0(a)\neq \hat{b}_{-0}(a)$).

All the representations of the harmonic oscillator corresponding to different frequencies are always related by a so called squeezing relation (see Ref. \cite{Kim2001}). The invariant and the diagonal representations are related by a Bogoliubov transformation, that is, for each single mode there is a relation such that
\be
\hat{b}_n = \mu^*_n \hat{c}_n + \nu_n \hat{c}_{-n}^\dag \, ,\qquad \,\, \hat{b}_n^\dag = \mu_n \hat{c}^\dag_n + \nu^*_n \hat{c}_{-n} ,
\ee 
where, from Eqs. (\ref{c1}-\ref{c2}) and (\ref{b1}-\ref{b2}), 
\beq\label{mu}
\mu_n^* &=& \frac{1}{2} \left( \f{1}{R} \sqrt{\f{1}{\cN_n \omega_n}} + R \sqrt{\cN_n \omega_n} - i \dot{R} \sqrt{\f{ \cN_n}{\omega_n}} \right) , \\ \label{nu}
\nu_n &=& \frac{1}{2} \left( \f{1}{R} \sqrt{\f{1}{\cN_n \omega_n}} - R \sqrt{\cN_n \omega_n} - i \dot{R} \sqrt{\f{ \cN_n}{\omega_n}} \right) ,
\eeq
with $|\mu_n|^2 - |\nu_n|^2 = 1$. The ground state of the invariant representation can then be written as \cite{Mukhanov2007}
\be\label{sop}
|0_{n,-n}\rangle_{(b)} = \frac{1}{|\mu_n|} \sum_{k=0}^\infty \left( \frac{\nu_n}{\mu_n} \right)^k |  k_n, k_{-n}\rangle_{(c)} .
\ee
It can be checked that the Lewis representation turns out to be the diagonal representation in the limit of large values of the scale factor. In that limit, $\mu_n \rightarrow 1$ and $\nu_n \rightarrow 0$, and the ground states that correspond to both representations turn out to be the same, i.e.,
\be
\lim_{a\rightarrow\infty}|0_{n,-n}\rangle_{(b)} =  |0_{n,-n}\rangle_{(c)} .
\ee

Let us summarize this section by discussing the interpretation of both representations we used. The most natural representation (in the sense that is more direct and standard) is the diagonal representation given by the $\hat{c}$ operators. However, the very fact that we are dealing with a system analogous to an harmonic oscillator in a dispersive medium, avoids the possibility that the number operator constructed with the diagonal representation be invariant. Taking into account that we are working under certain sensible boundary conditions, given by the independence of the global properties of the multiverse with respect to the scale factor of a single universe, we were forced to look for another representation, the Lewis representation, whose number operator remains invariant.

Thus, the Lewis representation is constructed in order to deal with the boundary conditions of the whole multiverse (as a collection of universes), so we can associate to it a notion of external observer. In the limit of large values of the scale factor, the Hamiltonian written in terms of the Lewis representations acquires a diagonal form (the Lewis and the diagonal representations coincide in this limit). Indeed, we expect that in this limit the universes become asymptotically independent. In this way, it is reasonable to relate the diagonal representation with an observer inside a given universe (that could be us). To conclude, notice that for each value of the scale factor we obtain both representations (Lewis and diagonal), but it is possible to relate all of them with a Bogoliubov transformation.


\section{Thermodynamics of entanglement in the multiverse}\label{section4}

We have described a multiverse scenario with a collection of universes created in pairs whose quantum states are given by the equation of an harmonic oscillator. In this section we compute the thermodynamic properties derived from the entanglement between the universes.

If we consider the Lewis representation, that is, the representation of an external observer, the operator $\hat{N}_{b,n}$ has the interpretation of the excitation number of the multiverse. Given that we are describing the whole multiverse, where by definition there is no external force, we expect the multiverse to stay in the ground state, at least as a first approximation. Then, in order to study the thermodynamics of entanglement, it is sensible to consider the ground state of two universes whose quantum states are given in the $b$-representation,  $|0_{n,-n}\rangle_{(b)}$, for a particular value of the field mode $|n|$. Nevertheless, the procedure followed here could be easily extended to more complicated states, or even general states, although the equations would be more intricate and it would be more difficult to obtain relevant results. In any case, we do not expect qualitative changes in the behavior of the thermodynamics for the excited states. Regarding the mode $n$ of the scalar field, we will see that the asymptotic limit for large values of $a$ is independent of the mode considered for the scalar field, although the subsequent corrections depend on it.

As we have shown in the preceding sections, the invariant and the diagonal representations are related by the Bogoliubov transformation (\ref{sop}). The density matrix $\rho$ that represents the quantum state of the entangled pair can be expressed as
\begin{equation}
\rho \equiv |0_{n,-n}\rangle_{(b)} \,_{(b)}\langle 0_{n,-n}| .
\end{equation}
It represents, in the invariant representation, the ground state of the entangled pair of universes that were born from the Euclidean instanton represented in Fig. \ref{fig1}. Following the same procedure of that used in the context of a quantum field theory in a curved background (see appendix \ref{app1}), the reduced density matrix $\rho_n$ (that represents the quantum state of one single universe of the entangled pair in the diagonal representation, more concretely, the universe with a positive value of the mode $n$) can be obtained by tracing out from $\rho$ the degrees of freedom of the partner universe. Using Eq. (\ref{sop}), it yields
\begin{equation}
\rho_n \equiv {\rm Tr}_{-n} \rho = \sum_{j=0}^\infty \langle j_{-n} | \rho | j_{-n} \rangle = \frac{1}{|\mu_n|^2} \sum_{j} \left( \frac{|\nu_n|}{|\mu_n|}\right)^{2 j} |j_n\rangle \langle j_n | 
\end{equation}
or, written in the Gibbs form
\begin{equation}\label{eq6325}
\rho_n(r) = \frac{1}{Z_n(r)} \sum_{j=0}^\infty e^{-\f{\omega_n}{T(r)}(j_n + \f{1}{2})} |j_n\rangle \langle j_n|,
\end{equation}
where $r$ is the parameter of squeezing, $|\mu_n| \equiv \cosh r$, $|\nu_n| \equiv \sinh r$, and  $Z_n^{-1} = 2 \sinh\frac{\omega_n}{2 T}$, with  $\f{\omega_n}{2}$ the energy that corresponds to the ground state of the positive modes in the diagonal representation, $|0_n\rangle_{(c)}$. The two universes of the entangled pair evolve then in thermal equilibrium with respect to each other, with a temperature 
\begin{equation}\label{eq6327}
T(r)=\f{\omega_n}{2\ln\f{1}{\tanh(r)}}
\end{equation}
that depends on the value of the parameter of squeezing $r$ (that in turn depends on the value of the scale factor). It is worth noticing that the thermal state represented by the density matrix (\ref{eq6325}) is indistinguishable from a classical mixture \cite{Partovi2008}. Thus, at first sight, the observers inside a single universe see their respective universes as classical universes. However, the quantum state (\ref{eq6325}) is the effect of partially tracing out the degrees of freedom of a partner universe in a composite entangled state which is a quantum state that has no classical counterpart \cite{Reid1986, PFGD1992a}.

The thermodynamic magnitudes of the thermal state given by Eq. (\ref{eq6325}) can easily be computed in terms of the squeezing parameter. For instance, the entanglement entropy \cite{Alicki2004, Vedral2006, Gemmer2009}
\begin{equation}
S_{ent}=-{\rm Tr}(\rho_{n}\ln\rho_{n}) ,
\end{equation}
turns out to be
\begin{equation}\label{eq6329}
S_{ent}(a) = \cosh^2 r \,\ln(\cosh^2 r) - \sinh^2 r \,\ln(\sinh^2 r) ,
\end{equation}
and the total energy associated to $\rho_n$   reads
\begin{equation}\label{eq49}
E_n(a) = {\rm Tr} (\rho_n \cH_n) = \omega_n (\sinh^2 r + \f12) ,
\end{equation}
where $\cH_n \equiv \omega_n (\hat{c}_n^\dag \hat{c}_n + \frac{1}{2})$. The change in the heat and work, as defined in Ref. \cite{Alicki2004, Gemmer2009}, are respectively
\begin{eqnarray}\label{eq50}
\delta W_n &=& {\rm Tr} (\rho_n \f{d \cH_n}{d a} ) = \dot{\omega}_n (\sinh^2 r + \f{1}{2}) , \\ \label{eq51}
\delta Q_n &=& {\rm Tr} (\f{d \rho_n}{d a} \cH_n ) = \omega_n \dot{r} \sinh 2r , 
\end{eqnarray}
from which it can be checked that the first principle of thermodynamics, $dE_n = \delta W_n + \delta Q_n$, is directly satisfied. From Eqs. (\ref{eq6329}) and (\ref{eq51}), it can also be checked that the production of entropy \cite{Alicki2004}, $\varsigma$, is zero 
\begin{equation}\label{eq6333}
\varsigma \equiv \f{d S_{ent}}{da} - \f{1}{T}\f{\delta Q_n}{d a} = 0 ,
\end{equation}
satisfying, therefore, the second principle of thermodynamics for any value of the scale factor. Furthermore, Eq.  (\ref{eq6333}) can be compared with the expression which is usually used to compute the energy of entanglement (see, Refs. \cite{Mukohyama1997, Mukohyama1998, Lee2007}),
\begin{equation}\label{eq53}
d E_{ent} = T dS_{ent} .
\end{equation}
It enhances us to establish an energy of entanglement given by
\begin{equation}
\label{eq86}
dE_{ent} = \delta Q_n  = \omega\sinh 2r \; dr .
\end{equation}

We can now compute explicitly the previous magnitudes. The first step is to compute the function $R(a,\vphi)$ that appears in the definition of $\hat{b}_{n}(a)$, Eqs. (\ref{b1}) and (\ref{b2}). The function $R$ is defined as $R(a,\varphi)= \sqrt{\phi_1^2(a) + \phi_2^2(a)}$, with  $\phi_1$ and $\phi_2$ two solutions of the Wheeler-DeWitt equation (\ref{GeneralWDW4}), which can be written, in the semiclassical limit, as
\begin{eqnarray}
\phi_1 &=& \sqrt{\f{1}{\cN_n(a)\,\omega_n(a,\vphi)}} \cos S_\omega(a,\vphi),\\
\phi_2 &=& \sqrt{\f{1}{\cN_n(a)\,\omega_n(a,\vphi)}} \sin S_\omega(a,\vphi) , 
\end{eqnarray}
so the expression for $R$ takes the form
\begin{equation}\label{eq54}
R \approx \sqrt{\f{1}{\cN_n(a)\,\omega_n(a,\vphi)}} .
\end{equation}
Then, inserting Eq. (\ref{eq54}) into Eqs. (\ref{mu}-\ref{nu}), we obtain for the entanglement parameters:
\begin{equation}
\mu \approx 1 + i \frac{\dot{R}}{2}\sqrt{\f{\cN_n}{\omega_n(a,\vphi)}},\qquad\quad
\nu \approx - i\f{\dot{R}}{2}\sqrt{\f{ \cN_n}{\omega_n(a,\vphi)}}.
\end{equation}
Therefore, the parameter of entanglement $r$ is given by
$$
\sinh r \equiv |\nu| = \left|\f{\dot{R}}{2}\sqrt{\f{ \cN_n }{\omega_n(a,\vphi)}}\,\right|.
$$
Taking into account that
$$
\dot{R}=-\f{R}{2}(\f{\dot{\cN}}{\cN}+\f{\dot{\omega}}{\omega})\sim - R |\Omega|\sin\theta,
$$
where $\theta$ was defined at the end of section \ref{section2} as the polar angle of the complex frequency $\Omega$,
we can easily compute the explicit value of $\sinh r$ at the leading order:
\be\label{sinhr1}
\sinh r=\f12 |\tan\theta|=\f12\left|\tan\left(\f12\arctan\left(\f{3\widetilde m\vphi^2}{H_0|1\mp 9\vphi^2/2|}\right)\right)\right|.
\ee
Some comments are in order at this point. First notice that, as we had anticipated, the parameter $r$ does not depend on the mode chosen at the leading order (although it will depend on it in the next order). Nevertheless, it crucially depends on the value of the cosmological constant $\Lambda_0$ as well as on the potential of the field $\vphi$. Particularly, in our case with $V(\varphi) = m^2 \varphi^2$, it depends on the value of $\vphi$ and on the mass $m$.

In order to compare our results with the case in which no scalar field is considered \cite{RP2012b, RP2012d}, it is interesting to compute the next two terms in the corrections of the expansion of $\sinh r$. It is a straightforward although a lengthy computation. If we consider the next order in the scale factor, $a$, in the expression for $\Omega$ (Eq. \ref{asintoticOmega}), that is,
\be 
\Omega_n^2 \sim \left( H_0^2 (1 \mp \f{9}{2}\varphi^2) \pm 3 i H_0 \widetilde m \varphi^2 \right) \sigma^2  a^4 +(2 \widetilde m\mp 3 i H_0) (n + \f12)  \, \sigma a ,\label{ecu60}
\ee
we obtain the following expansion for $\sinh r$ up to order $(1/a)^{4}$:
\be 
\sinh r = \left|f(\vphi)+g_n(\vphi)\f{1}{a^{3}}+O\left(\frac{1}{a^4}\right)\right|,
\ee
where $f(\vphi)=\f12|\tan\theta|$, as expected from Eq. (\ref{sinhr1}) and, in the limit of small values of the scalar field,  
\be
f(\varphi) \sim  -\f{3 \tilde{m}}{4 H_0} \varphi^2 \,, \qquad g_n(\varphi) \sim  \f{3(2n-1)}{8 H_0 \sigma}+\left(\f{27n}{8 H_0 \sigma}+\f{3\tilde{m}^2 (n+\f12)}{2 H_0^3}\right)\varphi^2.
\ee 

The consideration of a scalar field in the formulation of the inter-universal entanglement has important and unexpected consequences. Let us first consider the scale factor $a$ and the scalar field $\varphi$ as independent variables. In the $(a,\varphi)$-space, consider the limit of $\varphi \rightarrow 0$ for a constant value of the scale factor. Together with the value $n=-1/2$, it is equivalent to consider no scalar field, and therefore that limit can be useful to compare our results with previous ones. In such limit, 
\be \label{eq63}
\left.\lim_{\vphi\rightarrow 0} (\sinh r)\right|_{n=-\f12}=\frac{3}{4 H_0 \sigma a^3} ,
\ee
that exactly coincides with the results found in previous works \cite{RP2012b, RP2012d}. 

\begin{figure}
\centering
\includegraphics[width=8cm]{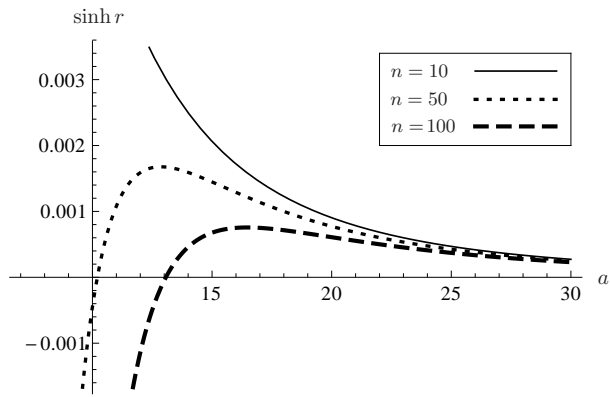}
\caption{We observe in this plot how the entanglement parameter (more concretely $\sinh r$) decreases for large values of the scale factor, $a$, for three different values of the mode $n$. Indeed, it decreases for all possible values of $n$.}
\label{grafica2}
\end{figure}

\begin{figure}
\centering
\includegraphics[width=8cm]{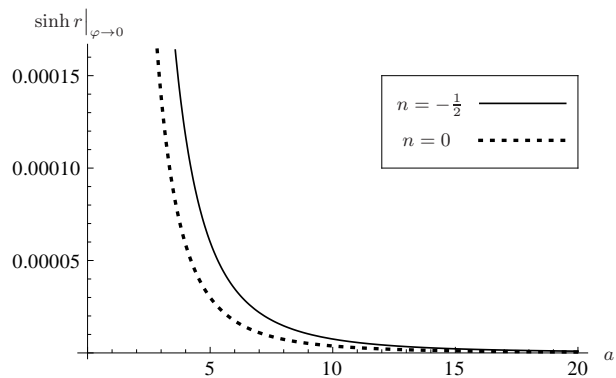}
\caption{In this plot we show the non zero contribution to the entanglement of the zeroth mode ($n=0$) of the scalar field, in comparison with the case where there is no field at all ($n+1/2=0$).}
\label{grafica1}
\end{figure}

During the oscillatory regime, after the slow-roll approximation has failed, the scalar field and the scale factor follow a classical trajectory in the $(a,\varphi)$-space. In particular, at late times the scale factor behaves like $a \sim e^{H t}$, and with the approximations made in this paper ($m \gg H$), the scalar field behaves like a dumped wave (see, for instance, Eq. (7.17) of Ref. \cite{Mukhanov2007}),
\be
\varphi_k(t) \sim e^{-\f32 H_0 t} e^{\pm i m t}  .
\ee
Then, the relation $\varphi(a)$ turns out to give $|\varphi(a)| \sim (a_0/a)^{\f32}$, where we have introduced a constant initial value for the scale factor, $a_0$, in order to be consistent with the dimensional analysis. This is a very interesting result because the leading order term is modified by the presence of the scalar field, and is given by
\be\label{eq66}
\sinh r \sim \left| \frac{3 (1-2n+2 a_0^3 \tilde{m} \sigma )}{8 H_0 \sigma  a^3} \right| .
\ee
This is a result coming from the consideration of the scalar field which introduces differences in the rate of entanglement between universes that depend on the values of the mode $n$ (see fig. \ref{grafica2}) and on the specific potential considered for the field, that is, on the value of $\tilde{m}$. Notice also that the quantum fluctuations of the vacuum state of the scalar field do contribute to the entanglement (see fig. \ref{grafica1}), as we have a different contribution to the entanglement with $\varphi\rightarrow 0$ for the case with $n=-1/2$ (considering no field from the beginning) or with $n=0$ (zeroth mode of the field). 

These results will have important consequences in the behavior of the thermodynamic magnitudes of entanglement of each single universe of the entangled pair, that we discuss in the following.

\subsection{Thermodynamic magnitudes of entanglement}

In this section we compute the thermodynamic magnitudes of entanglement associated to the state (\ref{eq6325}), which represents the quantum state of each single universe of an entangled pair. The temperature (\ref{eq6327}) turns out to be, at the leading order,
\be
T(\vphi,a)=\f{\xi(\vphi) \cos\theta}{2\ln\left(\f{\sqrt{1+\sinh^2(r)}}
{\sinh(r)}\right)}\:a^2\, ,
\ee
where $|\Omega| = \xi(\vphi)  a^2$, with $\xi(\vphi) \equiv \left( (H_0^2\mp\f{9H_0^2}{2}\vphi^2)^2+9H_0^2\widetilde m^2\vphi^4 \right)^\f{1}{4}$. It grows, essentially, because the frequency grows due to the expansion of the universe. However, the specific temperature per frequency $T/\omega$, is a measure of the rate of entanglement and it decreases as the universes expand, becoming asymptotically independent universes.

The standard measure of the entanglement between the parts of a composite system is the entropy of entanglement \cite{Vedral2006, Gemmer2009}. Nevertheless, both $\sinh r$ and the entropy $S$ behave in the same way, due to the fact that the entropy is a monotonic function of $\sinh r$ (given that $\sinh r \geq 0$). In the case being considered, the entropy of entanglement is expressed in Eq. (\ref{eq6329}) in terms of the parameter of entanglement $r$, which in the oscillatory regime is given by Eq. (\ref{eq66}). The entropy of entanglement turns out to be a monotonic decreasing function whose variation with respect to the scale factor behaves like
\be
\dot{S}_{ent} \sim - a^{-7} \log a .
\ee
It may still provide us with an arrow of time for each single universe of the entangled pair \cite{RP2012} because of its monotonicity. Let us also notice that the second principle of thermodynamics, which can be formulated as \cite{Alicki2004} the requirement of a non-negativity value of the production of entropy (\ref{eq6333}), is satisfied because the entanglement between universes is not an adiabatic process, in the quantum information sense, and the production of entropy (\ref{eq6333}) is zero for any change rate of the scale factor. Thus, the classical formulation of the second principle of thermodynamics applied to the entropy of inter-universal entanglement would provide us with no direction of time because $\dot{\varsigma}=0$ both for a decreasing and a increasing value of the scale factor. However, the quantum information formulation that states \cite{Plenio1998} that the entanglement rate between the parts of a composite system cannot be increased by any local operation and classical communication alone, does imply an arrow of time in each single universe of the entangled multiverse. By local operations we mean, in the context of the multiverse, processes that happen within each single universe of the multiverse. Therefore, anything that happens in a single real universe would make the entanglement rate between universes decrease as the universes expand. Besides, the entanglement arrow of time in the multiverse might be a testable property of each single universe provided that the entropy of entanglement is eventually related to the total entropy of the universe.

Regarding the energy, there are two contributions to the total energy given by the quantum information work and heat, $W$ and $Q$, respectively, in Eqs. (\ref{eq49}-\ref{eq51}). By inspecting these equations, we can easily see that the work $W$ is due to the variation of the frequency that is caused by the expansion of the universe. It is therefore the change of energy caused by the change of the proper volume of the universe. On the other hand, the variation of heat, Eq. (\ref{eq51}), is due to the change of the rate of entanglement, $r$. It is therefore an energy purely associated to the entanglement between universes and it supplies a correction term to the total energy of an unentangled universe. Thus, entangled and unentangled universes behave differently, so this effect makes inter-universal entanglement to be a falsifiable property of the universe in the multiverse scenario.

The energy of entanglement is usually defined \cite{Mukohyama1997, Mukohyama1998, Lee2007} as it is defined in Eq. (\ref{eq53}). Therefore, we can compute it by integrating Eq. (\ref{eq86}). First, we can check that the limiting value of a vanishing scalar field with $n=-1/2$ yields the same result to that obtained with no scalar field \cite{RP2012b,RP2012d}, as expected. For a large value of the scale factor, making use of equations (\ref{ecu60}) and (\ref{eq63}), we obtain for the energy of entanglement:
\be\label{ee1}
E_{ent} \equiv Q = \int da \, \omega_n \sinh 2r \, \dot{r} \sim \frac{27}{32  H_0 \sigma a^4}.
\ee
The consideration of the scalar field introduces a different behavior of the energy of entanglement. During the oscillatory phase of the scalar field, $|\varphi(a)|\propto (a_0/a)^{\f32}$, so using Eq. (\ref{eq66}) the energy of entanglement can be approximated for large values of the scale factor as
\be\label{ee2}
E_{ent} \sim Q \propto \frac{27 (1-2n+2 a_0^3 \tilde{m} \sigma)^2 }{128 H_0 \sigma  a^4} .
\ee
In both cases the energy of entanglement asymptotically vanish for an infinite value of the scale factor as it does the entanglement rate between the universes. Particularly, Eq. (\ref{ee2}) becomes Eq. (\ref{ee1}) for the limiting values $n \rightarrow -\f12$ and $\tilde{m}\rightarrow 0$, as it was expected. However, Eqs. (\ref{ee1}) and (\ref{ee2}) give different contributions to the total energy of the universe. That should have, in principle, observable consequences provided that the energy of inter-universal entanglement gives a contribution to the total energy of each single universe. 

Let us also notice two other consequences of Eq. (\ref{ee2}). First, as we already pointed out in the case of the  $\sinh r$, there is a non-vanishing contribution of the $n=0$ mode of the scalar field, that is, the quantum fluctuation of the vacuum state of the scalar field contribute. Second, the energy of entanglement, Eq. (\ref{ee2}), can be written as the energy of entanglement (\ref{ee1}) with an effective value of the Hubble parameter given by
\be
H_0^{eff} \equiv \frac{4 H_0}{\left(1-2 n+2 a_0^3 \tilde{m} \sigma \right)^2}.
\ee
We observe that for most of the values of the field mode (except those such that $n \sim a_0^3\tilde{m} \sigma$), the effective Hubble parameter satisfy $H_0^{eff}\ll H_0$. Therefore, we can conclude that, in general (for all the values of the field mode except the highly implausible case pointed out before), the effect of the scalar field in the value of the inter-universal energy of entanglement is equivalent to reduce the value of the effective cosmological constant, $\Lambda^{eff} \equiv 3 (H_0^{eff})^2 \ll \Lambda_0$.


\section{Conclusions}\label{conclusions}

The multiverse has become during the last years an interesting and fruitful scenario where exploring customary problems of classical and quantum cosmology. The third quantization procedure discussed here allows us to mimic, in the context of the entangled multiverse, well known techniques   used in quantum field theories in curved space-times which uncover fundamental phenomena of the space-times with event horizons, like those involved in the formulation of Hawking radiation or in the particle creation in a de-Sitter space-time. In both cases it is possible to relate different representations of the time-dependent harmonic oscillator by Bogoliubov transformations. We dealt here, however, with a new scenario in which two or more universes are entangled and we computed some of their thermodynamic properties of entanglement.

More precisely, our model of the multiverse consists in a collection of universes filled with a scalar field with a quadratic potential. The addition of such a scalar field provides a realistic framework that allows the study of both the inflationary stage of the universe and the subsequent oscillatory regime of the scalar field that would give rise to the current universe at large values of the scale factor. It is worth noticing that a similar formalism can be followed using other potentials, including those representing interacting matter fields, that would lead to the same equation (\ref{Schrodinger2}) with another corresponding Hamiltonian. Therefore, the developments made in this paper are rather general and can straightforwardly be applied to multiple cosmological scenarios.

In order to work out the model we have imposed certain boundary conditions. It is reasonable to assume that the global properties of the multiverse do not depend on the value of the scale factor of a specific universe. This is analogous to imposing space-time invariance to the vacuum state of a specific curved space-time, like it happens for instance in the choice of the vacuum state in a de-Sitter space-time \cite{Lapedes1978, Birrell1982, Mukhanov2007}. The parallel reasoning in the minisuperspace is to consider an invariant representation with respect to the value of the scale factor, which is formally the time-like variable. That makes the vacuum state of the multiverse to be invariant under reparametrizations of the scale factor and consequently under time reparametrizations in a particular single universe, as it is expected.

However, as it also happens in a de-Sitter space-time, in which an observer is better described by static coordinates, or in the case of the black hole radiation where the particles are meaningfully defined in the asymptotic flat region of the space-time, the representation of universes from the point of view of internal (actual) observers is preferably  given by the asymptotic representation of large universes with a large value of their scale factors. Then, continuing with the analogy with a curved space-time, the fact that a particular observer has no access to a region of the space-time (the region behind the event horizon in a de-Sitter or a Schwarzschild space-time), that is, the partner universe of the entangled pair, makes our universe to effectively stay in a thermal state. It is worth pointing out that for an internal observer such a thermal state of the universe is indistinguishable from a classical mixture \cite{Partovi2008, RP2012}. However, it comes from a composite entangled state that has no classical counterpart whatsoever \cite{Reid1986, PFGD1992a}.

The former boundary condition translates into the condition that the number operator associated with a given representation has to be invariant. We achieved it by using the Lewis representation defined by the operators $\hat{b}$ and $\hat{b}^\dag$. On the other hand, we considered the natural representation that diagonalizes the Hamiltonian and that represents large parent universes, plausibly with observers inhabiting them. The fact that both representations coincide in the asymptotic limit and given that the Lewis representation encodes global properties of the multiverse, suggest us that we can associate the invariant representation to an external observer, i.e., a hypothetical observer that would live in the multiverse.

In such a framework, we considered that the universes could be created in entangled pairs. Then, if we consider that the multiverse is in the ground state of the Lewis representation, an internal observer would see her universe in a state with a non-zero value of its excitation number. From our point of view, the consideration that the multiverse is in the ground state (in the Lewis representation) is sensible, and it is the way we followed in the present paper. Nevertheless, it would be possible to extend our results to a general quantum state, although the technical difficulties would increase and we do not expect qualitative differences, at least at the leading order of the asymptotic limit.

In this paper, we have generalized previous results on the thermodynamics of entanglement between universes by adding a scalar field. In this way, we enriched the model and considered a more realistic scenario. We obtained the thermodynamic magnitudes of entanglement for each single universe of an entangled pair of universes that generalize previous works on the subject \cite{RP2012b, RP2012d} recovering their results in the appropriate limits. The entropy of entanglement still supplies us with an arrow of time for each single universe. The energy of entanglement, given by the quantum information heat, provides us as well with a correction to the total energy of the expanding universe that would not be present in the case of an isolated universe, becoming thus a falsifiable property of the universe. In a pair of entangled universes, the energy of entanglement decays at late times becoming asymptotically zero for large values of the scale factor, where the universes become uncorrelated. The quantum fluctuations of the vacuum state of the scalar field contribute to the energy of entanglement between different universes. However, the more remarkable effect of the scalar field in the energy of entanglement would be the effective reduction of the cosmological constant in each single universe.

The contribution of the energy of entanglement to the total energy of each single universe would thus presumably have observable consequences on the dynamical properties of the universe, making therefore testable the whole multiverse proposal. However, the study of the dynamical consequences on a single universe of its entanglement properties with respect to other universes of the multiverse needs first of a framework that would account for the backreaction of the thermodynamic magnitudes of inter-universal entanglement on the physical properties of each single universe, which we expect to develop in future works.

In that sense, it is worth pointing out that although the complete relationship between the thermodynamic magnitudes of entanglement and those of the customary formulation of thermodynamics is not clear yet,  it is currently a promising area of study \cite{Plenio1998, Vedral2002, Amico2008, Anders2007, Brandao2008}. If such an ambitious program is finally achieved in quantum optics and quantum information theory, it will entail an important tool for testing the entanglement among universes, the multiverse proposal and the theories that underlie it. Therefore, in our opinion, the entanglement between universes represents a new interesting way to explore possible observable consequences of the multiverse, offering physically admissible models and a novel way to look at some of the main open problems in cosmology.

\section*{Acknowledgments}

This work was in part supported by the Spanish MICINN research
grants FIS2009-11893 and FIS2012-34379. IG is supported by the CNPq-Brasil. SRP was supported by the Basque Government project IT-221-07. 

\appendix

\section{Quantization of a scalar field in a de-Sitter space-time}\label{app1}

In this appendix we review the basic formalism for the quantization of a scalar field in the background of a curved space-time \cite{Parker1969, Lapedes1978, Birrell1982, Mukhanov2007}. The aim is twofold: on the one hand, we want to stress the formal resemblance between the third quantization formalism used in this paper and the quantization of fields in a curved space-time. On the other hand, we show that the formalism develop in section II.C naturally leads, starting from Eq. (\ref{hamiltonianh}), to the customary quantum field theory of a scalar field in a de-Sitter space-time.

The wave equation for a scalar field with mass $m$ that propagates in the background of a de-Sitter space-time is given by \cite{Birrell1982, Mukhanov2007}
\be
\chi'' - \Delta \chi + (m^2 a^2 - \frac{a''}{a}) \chi = 0 ,
\ee
where $\chi\equiv a \varphi$, $f'\equiv \frac{\partial f}{\partial \eta}$, with $\eta \equiv \int \frac{dt}{a}$ being the conformal time, and $\Delta$ is the Laplacian on the spatial sections of the space-time. The general solution is given by
\be
\chi(\vec{x},\eta) = \int \frac{d^3\vec{k}}{(2\pi)^\frac{3}{2}} \, \chi_{\vec{k}}(\eta) \, e^{i  \vec{k} \eta} ,
\ee 
where the amplitudes $\chi_{\vec{k}}(\eta)$ satisfy the equation of a harmonic oscillator, 
\be\label{ho}
\chi_{\vec{k}}'' + \omega^2_{\vec{k}}(\eta) \chi_{\vec{k}} = 0 ,
\ee
with a time dependent frequency given by
\be
\omega^2_{\vec{k}}(\eta) = |\vec{k}|^2 + m^2 a^2 - \frac{a''}{a} .
\ee
Different representations can be used for expressing the solutions of the harmonic oscillator (\ref{ho}). Let us consider two of those representations, with creation and annihilation operators $(\hat{a}^\dag, \hat{a})$ and $(\hat{b}^\dag,\hat{b})$, respectively, related by a Bogoliubov transformation \cite{Mukhanov2007}
\be
\hat{a}_k = \mu^*_k \hat{b}_k + \nu_k \hat{b}^\dag_{-k} \;\; , \;\; \hat{a}_k^\dag = \mu_k \hat{b}^\dag_k + \nu_k^* \hat{b}_{-k} ,
\ee
with $|\mu_k|^2 - |\nu_k|^2 =1$. Then, for a particular value $k$ of the mode the corresponding states are related by  \cite{Mukhanov2007}
\be
|0_{k,-k}\rangle_{(\hat{b})} = \frac{1}{|\mu_k|} \sum_{n=0}^\infty \left( \frac{\nu_k}{\mu_k} \right)^n |  n_k, n_{-k}\rangle_{(\hat{a})}.
\ee
That is, because of the isotropy the particles are produced in pairs with opposite momenta $\vec{k}$ and $-\vec{k}$. If for any reason we would have access only to the positive modes of the field, then the reduced density matrix that represents the quantum state of the scalar field would be given, in the $a$-mode representation as
\be
\rho_k \equiv  {\rm Tr}_{-k} |0_{k,-k}\rangle_{(\hat{b})} \langle 0_{k,-k}|= \sum_{j=0}^\infty \frac{1}{|\mu_k|^2} \sum_{n,m=0}^\infty \frac{\nu_k^n (\nu_k^*)^m}{\mu_k^n (\mu_k^*)^m} \langle j_{-k} | n_k, n_{-k}\rangle \langle m_k, m_{-k}|j_{-k}\rangle = \frac{1}{|\mu_k|^2} \sum_{n} \left( \frac{|\nu_k|}{|\mu_k|}\right)^{2 n} |n_k\rangle \langle n_k |,
\ee
which is a Gibbs state that represents a thermal state with temperature $T$ given by
\be
T = \frac{\omega_k}{2 \log\frac{|\nu_k|}{|\mu_k|}} \left( \frac{\hbar}{k_B} \right)  .
\ee
From this expression for the temperature we can easily compute the customary thermodynamic magnitudes.

Let us stress two remarkable things. First, the $k=0$ mode of Eq. (\ref{ho}) is also obtained from the Heisenberg equations
\be
\f{\partial \varphi}{\partial t} = i [ \hh , \varphi] \,,\quad \,\,\, \f{\partial p_\varphi}{\partial t} = i [ \hh , p_\varphi] ,
\ee
with $\hh$ being given by Eq. (\ref{hamiltonianh}). Thus, Eq. (\ref{hamiltonianh}) naturally leads to the customary quantization of a scalar field with mass $m$ in a de-Sitter space-time provided that we just consider a homogeneous and isotropic scalar field, $\varphi(\vec{x}, t) \equiv \varphi(t)$. It is straightforward to show \cite{Mukhanov2007} that for other modes different from zero the hamiltonian is given by
\be
\hh \equiv \frac{1}{2 M(t)} p_\vphi^2 + \f{M(t)}{2} \left( \f{1}{a^2} (\nabla \varphi)^2 + m^2 \varphi^2 \right) .
\ee

Secondly, Eq. (\ref{ho}) is formally similar to Eq. (\ref{GeneralWDW1}), where the scale factor formally plays the role of the time variable and it has been used the covariant generalization of the Laplacian operator in the minisuperspace, given by Eq. (\ref{LB}). This fact is at the heart of the parallelism made throughout this paper between the third quantization of the wave function of the universe and the generalization of a scalar field in a curved background.


\end{document}